\documentclass[pra, twocolumn, superscriptaddress, nofootinbib, showpacs]{revtex4}
\usepackage{amsmath}
\usepackage{amsfonts, amssymb}
\usepackage{dsfont}
\usepackage{graphicx}
\usepackage{hyperref}
\usepackage{xcolor}

\newcommand{\ket}[1]{\left| #1 \right>}

\begin{document}

\title{Adiabaticity and diabaticity in strong-field ionization}

\author{Antonia Karamatskou}
\email{antonia.karamatskou@cfel.de}
\affiliation{Center for Free-Electron Laser Science, DESY, D-22607 Hamburg, Germany}
\affiliation{Department of Physics, University of Hamburg, D-20355 Hamburg, Germany}

\author{Stefan Pabst}
\affiliation{Center for Free-Electron Laser Science, DESY, D-22607 Hamburg, Germany}
\affiliation{Department of Physics, University of Hamburg, D-20355 Hamburg, Germany}

\author{Robin Santra}
\email{robin.santra@cfel.de}
\affiliation{Center for Free-Electron Laser Science, DESY, D-22607 Hamburg, Germany}
\affiliation{Department of Physics, University of Hamburg, D-20355 Hamburg, Germany}


\date{\today}

\begin{abstract}
If the photon energy is much less than the electron binding energy, ionization of an atom by a strong optical field is often described 
in terms of electron tunneling through the potential barrier resulting from the superposition of the atomic potential and the potential 
associated with the instantaneous electric component of the optical field. In the strict tunneling regime, the electron response to the 
optical field is said to be adiabatic, and nonadiabatic effects are assumed to be negligible. Here, we investigate to what degree this 
terminology is consistent with a language based on the so-called adiabatic representation. This representation is commonly used in various 
fields of physics. For electronically bound states, the adiabatic representation yields discrete potential energy curves that are connected 
by nonadiabatic transitions. When applying the adiabatic representation to optical strong-field ionization, a conceptual challenge is 
that the eigenstates of the instantaneous Hamiltonian form a continuum; i.e., there are no discrete adiabatic states. This difficulty 
can be overcome by applying an analytic-continuation technique. In this way, we obtain a rigorous 
classification of adiabatic states and a clear characterization of (non)adiabatic and (non)diabatic ionization dynamics. 
Moreover, we distinguish two different regimes within tunneling ionization and explain the dependence of the ionization probability on the pulse envelope. 
\end{abstract}

\maketitle


\section{Introduction}\label{intro}

The realm of strong-field physics has become a focal point of interest in the atomic, molecular, and optical physics community over the last two decades. 
This was particularly supported by the rapid development of lasers producing high intensities ($10^{14}-10^{15}$~W/cm$^2$) that generate forces comparable to intra-atomic forces, and ultrashort pulse durations of the order of femtoseconds ($10^{-15}$~s) down to attoseconds ($10^{-18}$~s) \cite{gil,kra}. 
The time-resolved investigation of electron dynamics in atoms and molecules has come into reach because the typical time scales involved in electronic excitations (between 50~as and 50~fs) can be accessed. 

The process of tunneling ionization has been studied extensively. Following the calculation of the tunneling ionization rate 
for the ground state of hydrogen in a static electric field by Landau \cite{lan}, Keldysh extended the theory to 
ionization by strong electromagnetic fields \cite{kel}. Later, Ammosov, Delone, and Krainov (``ADK'') generalized 
the results to slowly varying fields by introducing the quasistatic approximation and defining the tunneling ionization rate 
by averaging over one optical period (``ADK theory'') \cite{adk}. A self-contained derivation of the tunneling rate 
in this approximation is presented in Ref.~\onlinecite{bis}.
In the original derivation \cite{kel} Keldysh introduced the parameter 
$
 \gamma= \sqrt{I_p/(2 U_p)},
$
which is now known as the Keldysh parameter \cite{iva}. Here, $I_p$ is the ionization potential and $U_p$ is the ponderomotive potential, 
which corresponds to the average energy of a free electron oscillating in the electric field.
According to Keldysh, $\gamma$ divides the phenomenon of strong-field ionization into two regimes: for $\gamma\ll 1$ 
ionization is governed by tunneling ionization \cite{lan}, while for $\gamma \gg 1$ the process is governed by perturbative multiphoton ionization \cite{mai}. 
In the range of $\gamma\approx 1$ both effects compete with each other \cite{pop, yam}. In later papers the Keldysh parameter has been connected to the notion of adiabaticity of the ionization process \cite{mev, bec}. Far into the tunneling regime, the atomic response is considered to be purely adiabatic. 
Adiabatic means in this context that the ionization rate at a given time is solely defined by the instantaneous electric field.

More generally, when a time-dependent process is adiabatic, the state of the system at any given time is always an eigenstate of the instantaneous Hamiltonian, which depends on one or more external parameters (like the electric field).
Consequently, the energy eigenstates and their corresponding eigenenergies become parametrized and lead to energy curves (or energy hyperplanes depending on the number of external parameters). 
Nonadiabatic dynamics occur when transitions between adiabatic curves start to appear. 
This is, particularly, the case when two adiabatic curves are energetically close to each other and the external parameters are changed relatively fast such that the system has no time to ``instantaneously'' respond to the change.
As a result, the system is not in one defined adiabatic state anymore but rather in a superposition of several adiabatic eigenstates. 
In various fields of physics and chemistry the adiabatic representation has been used to study adiabatic and nonadiabatic effects.
Its application includes fields like Rydberg atoms \cite{rub, cla}, molecular dynamics \cite{sti, tul, sch}, atomic and molecular collisions \cite{pec, smi, mil}, and ultracold gases and trapped ions \cite{blo, due, ste}.

An important aspect in the adiabatic representation is the discreteness of eigenstates which is essential to obtain a discrete set of energy curves. 
In the case of strong-field ionization, however, the instantaneous eigenstates of the Hamiltonian form a continuum.
Therefore, the identification of a nonadiabatic effect happens rather indirectly \cite{pot, zhe}: either the spectrum of the photoelectron 
after the pulse or the field dependence of the ionization rate is analyzed. 
Various results on nonadiabatic behavior in strong-field ionization have been presented in the literature \cite{arm, wan, yud} 
and there are many different usages of the terms ``adiabatic'' and ``nonadiabatic''.
By introducing an analytic continuation in the complex plane the instantaneous Hamiltonian becomes non-Hermitian and tunneling states appear as {\em discrete} eigenstates.
These discrete states can be now used to apply the adiabatic representation to strong-field ionization dynamics.

In this paper we strictly apply the adiabatic representation to strong-field ionization and find that in the tunneling regime the ionization dynamics is defined by a {\em diabatic} rather than an adiabatic behavior. Diabatic dynamics means that the response of the system follows one specific diabatic state.
Here, the diabatic states are defined by the overlap with the field-free eigenstates. 
In this formulation we find that the ionization dynamics can be divided into two regimes. Furthermore, with increasing frequency we observe a transition from the {\em diabatic} to the {\em nondiabatic} regime.
In particular, we study the few-cycle limit and find a non-constant population as a function of the optical frequency which has been interpreted in the literature as a sign of a nonadiabatic process \cite{bec, zhe}. 
We show for a few-cycle pulse with a Keldysh parameter $\gamma\ll 1$ that this effect rather represents a dependence on the form of the pulse and can be fully explained by a diabatic picture depending on a single diabatic state connected to the field-free ground state. 
The main text is divided into three sections:
\begin{itemize}
 \item Section~\ref{adiabatic} is devoted to the general theory of the equations of motion in the adiabatic basis and introduces also diabatic states.
 \item The third section presents one-photon absorption as an extreme case of a nonadiabatic/nondiabatic ionization process.
 \item In Sec.~\ref{strong}, the central section of this paper, we develop the concept of diabaticity in strong-field ionization.
       We examine the transition from the diabatic to the nondiabatic ionization regime.
\end{itemize}
Atomic units are employed throughout unless otherwise indicated.


\section{Adiabatic eigenstates}\label{adiabatic}
Whenever a system is given time to adjust to the parameters on which it depends, the response is called adiabatic.
In the following, we derive the quantum-mechanical equations of motion in the adiabatic basis, which is given by the states that are eigensolutions 
to the Hamiltonian of the system for a set of instantaneous parameters.

Let us study a system where the Hamiltonian depends on an external time-dependent parameter $ \epsilon(t)$. 
The time-dependent Schr\"odinger equation has the form
\begin{equation}
 i\partial_t | \Psi (t)\rangle = \hat{H}(t) | \Psi(t) \rangle = \left\{ \hat{H}_0 + \hat{U}[\epsilon(t)]\right\} | \Psi(t) \rangle. \label{tdse}
\end{equation}
$\hat{H}_0$ describes the atomic Hamiltonian, whereas $\hat{U}$ includes all external potentials and is dependent on the parameter $\epsilon(t)$. 
At a given time $t$, the instantaneous eigenstates, which constitute the adiabatic basis, are defined by\footnote{The time dependence is implicit via the parameter $\epsilon(t)$.}
\begin{equation}
\left[ \hat{H}_0 + \hat{U}(t) \right] |\Psi _n(t)\rangle=E_n (t)|\Psi _n(t)\rangle. \label{adbas}
\end{equation}
To analyze adiabatic and nonadiabatic effects we expand the electronic wavefunction in terms of the adiabatic eigenstates, $ | \Psi (t)\rangle = \sum_n \alpha_n(t)  |\Psi _n(t)\rangle$. 
Upon inserting this expression into Eq.~(\ref{tdse}) and projecting onto the eigenstate 
$|\Psi_m(t)\rangle$, the equation of motion for the coefficient $\alpha_m(t)$ reads
\begin{equation}
i \dot {\alpha} _m(t)+i \sum_n\alpha_n(t) \langle \Psi_m(t)|\partial_t|\Psi_n(t)\rangle
=\alpha_m(t) E_m(t). \label{eom}
\end{equation}
The off-diagonal matrix elements $\langle \Psi_m(t)|\partial_t|\Psi_n(t)\rangle$ introduce couplings between different adiabatic eigenstates, thus making the dynamics nonadiabatic~\cite{zen}. 
In the adiabatic approximation, where these couplings are considered to be very small, Eq. (\ref{eom}) becomes
\begin{equation}
i \dot {\alpha} _m(t)+i \alpha_m(t) \langle \Psi_m(t)|\dot{\Psi}_m(t)\rangle=\alpha_m(t) E_m(t),
\end{equation}
which is solved with the initial condition $\alpha_m(0) = 1$ by
\begin{equation}
\alpha_m(t) =\exp\bigg[-i\int_0^t dt' E_m(t')\bigg]\exp[i \gamma_m(t)],\label{coeff}
\end{equation}
where $\gamma_m(t)=i\int_0^t dt'\langle \Psi_m(t')|\dot{\Psi}_m(t')\rangle$, so that the system evolves in a specific adiabatic eigenstate with a phase. If, on the other hand, 
$\langle \Psi_m(t)|\dot{\Psi}_n(t)\rangle $ cannot be neglected, the whole sum in Eq.~(\ref{eom}) 
has to be considered, so that different adiabatic eigenstates get coupled and nonadiabatic motion emerges. 
We can use $\partial_t= \frac{\partial \epsilon}{\partial t}\partial_{\epsilon}$ 
and express the off-diagonal coupling elements also in terms of the change in $\epsilon$: 
\begin{equation}
 \langle \Psi_m|\dot{\Psi}_n\rangle =\langle \Psi_m|\partial_{\epsilon}{\Psi}_n\rangle  
\frac{\partial \epsilon}{\partial t}.\label{nonadcoup}
\end{equation}

\begin{figure}[htbp]
 \centering
 \includegraphics[width=\linewidth]{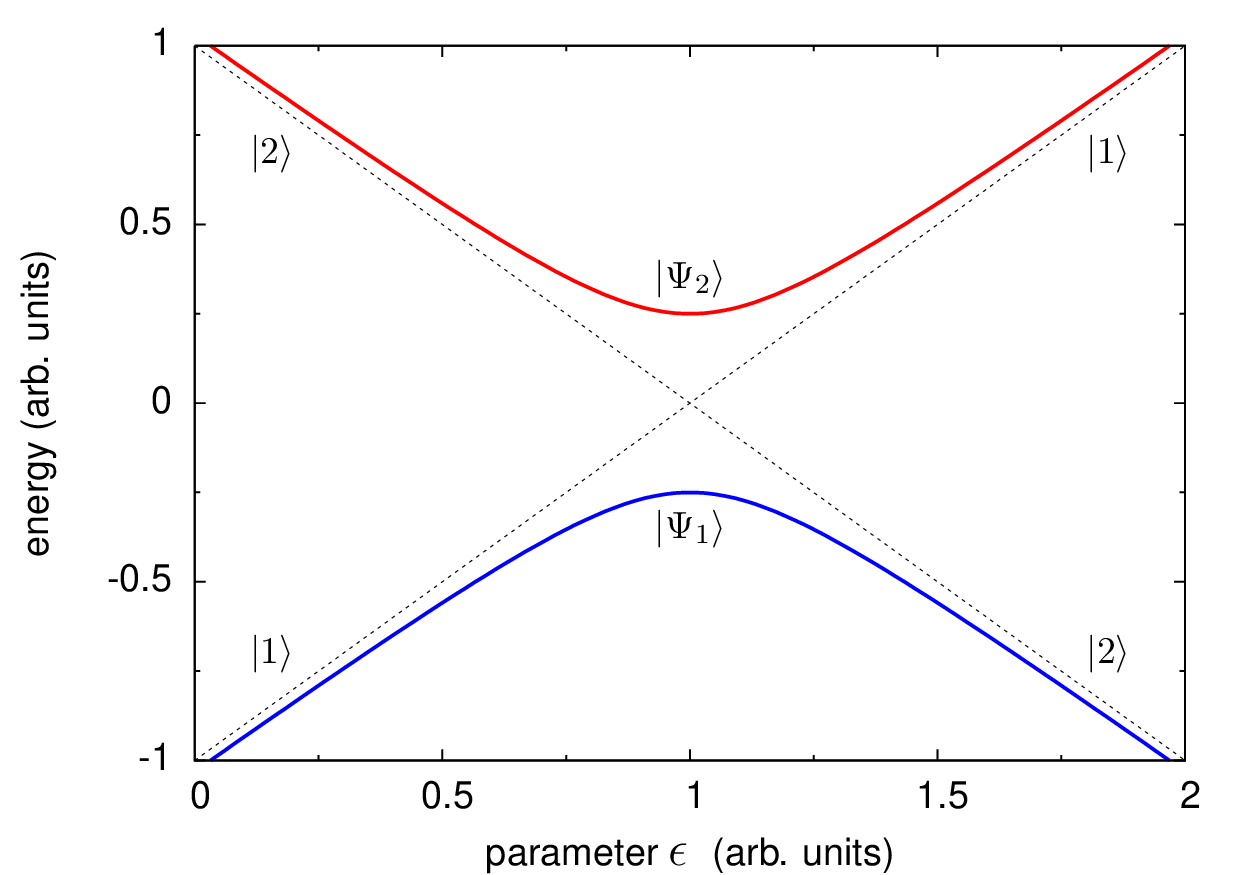}
 \caption{The energy curves of the two adiabatic states $|\Psi_1\rangle$ and $|\Psi_2\rangle$ are shown as functions of the parameter $\epsilon$. 
Through the off-diagonal matrix element $\Delta/2$ a nonadiabatic transition is possible, whereupon the system follows the diabatic states $\ket{1}$ and $\ket{2}$, respectively.}
 \label{fig0}
\end{figure}

Considering a two-level system with an external perturbation proportional to $\epsilon$, the Hamiltonian of the system takes the form
\begin{equation}
\hat{H} = 
\underbrace{
\begin{pmatrix} 
 -1 & 0 \\
  0 & 1
\end{pmatrix}+
\frac{\Delta}{2}
\begin{pmatrix} 
0 & 1 \\
1 & 0
 \end{pmatrix}}_{\hat{H}_0}+
\underbrace{
\frac{\epsilon}{2}
\begin{pmatrix} 
 1 & 0 \\
 0 & -1
 \end{pmatrix}}_{\hat{U}}
,
\end{equation}
where $\Delta$ is an internal coupling parameter. In Fig.~\ref{fig0} the energy curves of the two adiabatic states $\ket{\Psi_1}$ and $\ket{\Psi_2}$ of  this system are shown as a function of the external parameter $\epsilon$, assuming $\Delta=1$.
The internal coupling between the diabatic states $\ket{1}= (1,0)^T$ and $\ket{2}=(0,1)^T$ results in the effect that the two adiabatic curves do not cross.
This phenomenon is known as an ``avoided crossing''. We see that $\Delta$ is the energy splitting between $\ket{\Psi_1}$ and $\ket{\Psi_2}$ at the degeneracy point of the states $\ket{1}$ and $\ket{2}$.
If the parameter $\epsilon$ is changed sufficiently slowly, the system will remain in a given adiabatic state $\ket{\Psi_i}$ if, for $\epsilon\ll1$ or $\epsilon\gg 1$, the system was in the state $\ket{\Psi_i}$.
Note that in the vicinity of $\epsilon=1$ the character of the adiabatic states changes from $\ket{1}$ to $\ket{2}$ and vice versa.
If $\epsilon$ changes rapidly in the vicinity of $\epsilon=1$, the system has no time to change the character of its state; it makes a transition from one adiabatic state to the other and follows the diabatic states $\ket{1}$ and $\ket{2}$, respectively.
These jumps between adiabatic curves make the resulting dynamics nonadiabatic.
For a given value of the external parameter, we can obtain the diabatic states also by choosing the adiabatic eigenstates with the maximal overlap with the free states ($\epsilon=0$). For $\epsilon<1$, the diabatic state $\ket{1}$ has the maximal overlap with the adiabatic state $\ket{\Psi_1}$, while for $\epsilon>1$ the overlap of state $\ket{1}$ with the adiabatic state $\ket{\Psi_2}$ is maximal, and vice versa for the diabatic state $\ket{2}$. Asymptotically, the states $\ket{1}$ and $\ket{2}$ correspond to the states $\ket{\Psi_1}$ and $\ket{\Psi_2}$ before the avoided crossing, and vice versa after the crossing. Near the crossing an interpolation is performed in order to obtain a continuous and smooth state.

A system's dynamics can of course also be formulated in other representations, e.g., in a diabatic basis \cite{lic}, where the diabatic states do cross (see the states $\ket{1}$ and $\ket{2}$ in Fig.~\ref{fig0}). Usually the basis is chosen such, that the off-diagonal couplings in Eq.~\eqref{nonadcoup} vanish or are at least small \cite{smi, bae}. However, the diabatic basis, which is derived from the adiabatic basis by a unitary transformation, is not unique and there are many different approaches for reaching a diabatic representation \cite{baer, thi, sad}. One practicable method of diabatization is a local diabatization method, which means that the diabatic state is constructed piecewise in a two-level model: At each avoided crossing between two adiabatic states the diabatic state is followed. To this end, the size of the overlap with the corresponding field-free state can be used as a criterion. This method turns out to be fruitful for the description of diabatic and nondiabatic strong-field ionization (see Sec.~\ref{strong}). Once a diabatic representation has been found, one can ask with which rate transitions between diabatic states occur. These transitions will be called nondiabatic. 

In the following section we will make use of the fact that for weak perturbations the adiabatic eigenstates can be approximated through the field-free eigenstates. Therefore, the diabatic states exhibiting the maximal overlap with the field-free states are also the adiabatic states. In this case, the nondiabatic transitions are exactly the nonadiabatic transitions described above.


\section{One-photon absorption}\label{onephot}
First, we analyze the case of one-photon absorption within the adiabatic representation.
If the system is exposed to a weak electric field of the form 
$F(t)=F_0 \cos(\omega t)$ (in the dipole approximation, see Sec.~\ref{strong}), with a frequency $\omega$, the system Hamiltonian is perturbed 
by the term $F(t)\,\hat z$ \cite{coh}, where $\hat z$ is pointing in the direction of the field (which is assumed to be linearly polarized). 
The Hamiltonian in Eq.~(\ref{tdse}) takes the form
\begin{equation}
 \hat{H}(t) = \hat{H}_0 + F(t)\hat{z}, \label{field}
\end{equation}
where $\hat{H}_0$ is the atomic Hamiltonian and the electric field $F(t)$ is coupled 
classically to the dipole operator $\hat{z}$ of the electron [the field $F(t)$ corresponds to the parameter $\epsilon$ of Sec.~\ref{adiabatic}]. 

In the following, we show that in the adiabatic representation the off-diagonal coupling elements in Eq.~(\ref{eom}) are crucial for introducing transitions. 
Let $\{\Psi_{n}^{(0)}\}_{n=0}^\infty$ be the eigenstates of the field-free Hamiltonian, $H_0 |\Psi_{n}^{(0)}\rangle 
= \omega_n |\Psi_{n}^{(0)}\rangle$. For simplicity, we assume that the initial and final states of interest in the one-photon transition are nondegenerate. 
Performing static perturbation theory to first order, the adiabatic eigenstates read~\cite{fri}
\begin{equation}
|\Psi_{n}^{(1)}\rangle
=
|\Psi_{n}^{(0)}\rangle+\sum_{k\neq n}\frac{\langle \Psi_{k}^{(0)}|F\hat{z}|\Psi_{n}^{(0)}\rangle}{\omega_n-\omega_k}|\Psi_{k}^{(0)}\rangle.
\label{expr}
\end{equation}
Inserting Eq.~(\ref{expr}) in Eq.~(\ref{nonadcoup}) with $\epsilon$ being the field $F$, we obtain the nonadiabatic coupling elements to first order in $F$:
\begin{equation}
 \frac{\partial F}{\partial t}
\langle\Psi_{m}^{(0)}|  \sum_{k\neq n}\frac{\langle \Psi_{k}^{(0)}|\hat{z}|\Psi_{n}^{(0)}\rangle}{\omega_n-\omega_k}|\Psi_{k}^{(0)}\rangle
=\frac{\partial F}{\partial t} \frac{\langle \Psi_{m}^{(0)}|\hat{z}|\Psi_{n}^{(0)}\rangle}{\omega_n-\omega_m}.
\label{firstorder}
\end{equation}
We are now ready to solve Eq.~\eqref{eom} including nonadiabatic coupling. We may treat the operator 
$\hat{V}_F~\!=~\!\frac{\partial F}{\partial t}~\partial_F$ 
as a perturbing time-dependent operator and, hence, analyze the states with time-dependent perturbation theory \cite{fri}. 
The first-order correction to the zeroth-order coefficient [Eq.~\eqref{coeff}] is given by
\begin{equation}
 \alpha_f^{(1)}(t) = -i\int_0^{t} dt' e^{i(\omega_f-\omega_i)t'}\frac{\partial F}{\partial t'}\frac{\langle\Psi_{f}^{(0)}|\hat{z}|\Psi_{i}^{(0)}\rangle}{\omega_f-\omega_i}.
\end{equation}
Assuming $\omega_f>\omega_i$, we obtain the total transition probability per unit time
\begin{equation}
 w_i=\sum_{f} \frac{|\alpha_f^{(1)}|^2}{t}=2\pi \sum_{f} \bigg|\langle \Psi_{f}^{(0)}|\frac{F_0 \hat{z}}{2}
|\Psi_{i}^{(0)}\rangle\bigg| ^2 \delta(\omega_f-\omega_i-\omega) .
\end{equation}
This equation is exactly Fermi's golden rule \cite{sak}. In the present approach it is the 
nonadiabatic coupling that induces one-photon transitions between the field-free eigenstates. 
Viewed in this way, the phenomenon of one-photon absorption is entirely nonadiabatic.
In the one-photon case, the states are well separated by a large energy gap and there is no avoided crossing 
due to the weak field, which is only a perturbation to the field-free states. 
Note that the adiabatic states coincide with the diabatic states in the weak-field limit.


\section{Strong-field ionization of atoms}\label{strong}
While in the case of one-photon ionization the photon energy necessarily exceeds the ionization potential, 
we will now examine the situation where the atomic system is irradiated 
by an intense electric field $F(t)$ with a low photon energy, i.e., many photons are needed to ionize the atom.

\begin{figure}[htbp]
\centering
\includegraphics[width=\linewidth]{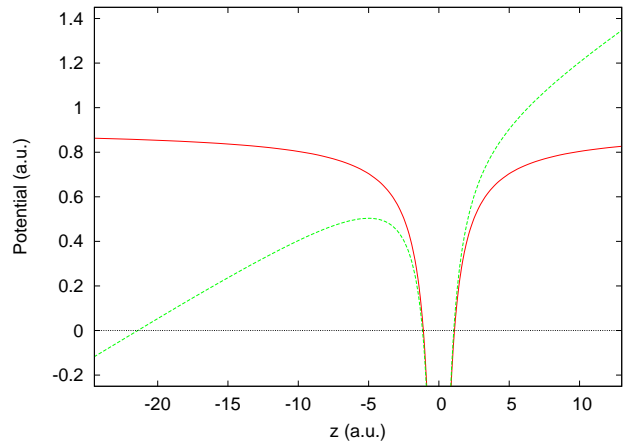}
\caption{The pure Coulomb potential of the helium atom (solid red line) is tilted in the presence 
of the electric field (dashed green line). The dotted black line denotes the field-free ground-state energy.}
\label{fig1}
\end{figure}

When applying a strong external field [see Eq.~(\ref{field})] the effective potential seen by an electron gets tilted (see Fig.~\ref{fig1}).
Therefore, a barrier of finite height is created through which the electron can tunnel. 
(If the electric field is so strong that the electron's energy lies above the barrier, 
the electron can just leave the atom without tunneling. This effect is called above-barrier ionization \cite{ebe, scr}.)
This tunneling picture of a tilted potential relies on the length form of the light-matter interaction, i.e., $F(t)\,\hat{z}$. 
Furthermore, the form of the Hamiltonian [cf. Eq.~\eqref{field}] is a result of the dipole approximation, which holds in our case, 
because the size of the system of interest (a few \AA) is much smaller than the wavelength of the light pulse ($\approx 1~\mu$m) \cite{lou, cra}. 

In order to describe strong-field ionization dynamics, the Schr\"odinger equation of the atom exposed to the field has to be solved nonperturbatively
because perturbation theory fails for these high field strengths. As shown in Sec.~\ref{adiabatic} in the adiabatic case the system will follow a given 
adiabatic state without making any transition. However, in the presence of a static electric field, electronically bound states become tunneling states, which means that there is ionization via tunneling. 

In the following, we will study helium as a concrete example to illustrate tunneling ionization within the framework of the adiabatic representation.


\subsection{Constructing adiabatic and diabatic states for helium}
As already discussed (see Sec.~\ref{intro}), in strong-field ionization the spectrum forms a continuum where a direct application of the adiabatic representation is inconvenient.
To overcome this problem, a rigorous analytical continuation of the Hamiltonian can be performed
by rotating the electron coordinates about an angle into the complex plane; this procedure is called complex scaling \cite{nim}. 
Another way to generate discrete eigenstates is to add a complex absorbing potential (CAP) to the Hamiltonian \cite{rismey}. 
It can be shown that the latter method, which is conceptually easier, is closely connected to the complex scaling approach \cite{ris}. 
The key idea here is that for every tunneling state, i.e., every adiabatic atomic state that allows the electron to tunnel 
through the field-induced barrier, there exists a discrete eigenstate --- a so-called Gamow vector \cite{boh} or Siegert state \cite{sie}
--- of the instantaneous Hamiltonian. A Siegert state is associated with a complex energy and lies outside the Hermitian domain of the Hamiltonian. 
In fact, the associated wavefunction is exponentially divergent for large distances from the atom. Complex scaling or the use of a
CAP eliminates the divergent behavior and renders the tunneling wavefunction square integrable. 
Thus, by making the Hamiltonian non-Hermitian, it becomes possible to calculate, within Hilbert space, the complex Siegert energies of tunneling states.
The imaginary part of the Siegert energy $E$ provides the tunneling rate 
$\Gamma$ of each Siegert state by the relation $\Gamma = -2\ {\rm Im} (E)$ \cite{moi, san}.

In order to obtain the instantaneous eigenstates we solve Eq.~(\ref{adbas}) with the Hamiltonian in Eq.~(\ref{field}) including a CAP.
This yields the adiabatic eigenstates and corresponding eigenenergies
of the atom shown in Fig.~\ref{fig2}. A more detailed description of the methods used is given in the Appendix~\ref{app}.
\begin{figure}[!ht]
\centering
\includegraphics[width=\linewidth]{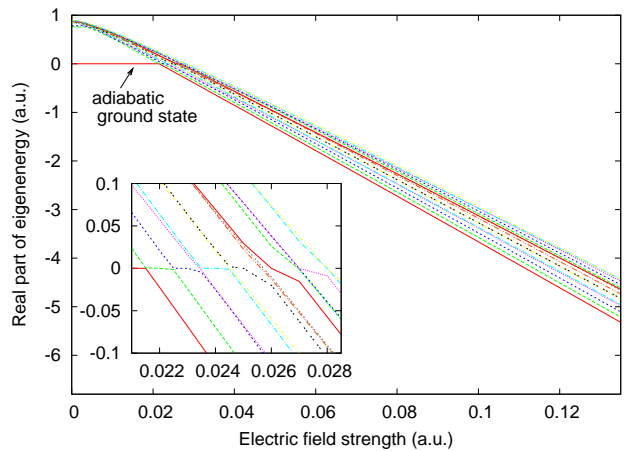}
\caption{The real part of the energy of the first adiabatic eigenstates as a function 
of a static electric field. The inset magnifies avoided crossings for small electric fields.}
\label{fig2}
\end{figure}
We observe many avoided crossings among the higher adiabatic eigenstates for field strengths 
in the range below $0.01$~a.u. ($1$~a.u.$=5.14\times 10^{9}$~V/cm), while the ground state energy does not change significantly. 
One might wonder whether for sufficiently slow ramping of the electric field the atom follows 
the adiabatic ground state. Indeed, for field strengths up to $0.02$~a.u. the adiabatic ground-state energy seems to remain constant. But we know  that the electric field can mix a whole manifold of excited states into the field-free states. When this happens, the adiabatic ground state loses the character of the field-free ground state (cf. Fig.~\ref{fig0}). Analyzing the avoided crossings involving the adiabatic ground state around the field strength of $0.02$~a.u., we find that the ramping of the field has to be so slow that it lies in the radio frequency regime. Therefore, the system does not follow the adiabatic ground state for the frequency range of light usually employed in experiments (typically around $800$~nm, corresponding to $4\times 10^{14}$~Hz). 

\begin{figure}[htbp]
\centering
\includegraphics[width= \linewidth]{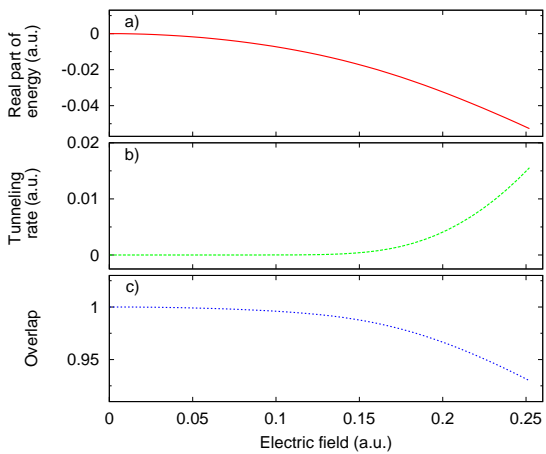}
\caption{(a) The real part of the energy of the diabatic state $|\Psi_0^{(d)}\rangle$, and (b) its tunneling rate, $\Gamma=-2{\rm Im}(E)$,
are shown as a function of the electric field. (c) The overlap of $|\Psi_0^{(d)}\rangle$ with the field-free ground state.}
\label{fig3}
\end{figure}

The electronic state 
follows the instantaneous eigenstate that has the maximal overlap with the field-free ground state. This is exactly the diabatic behavior described in Sec.~\ref{adiabatic}, where the electronic state jumps from one adiabatic state to the other, keeping its field-free character. Here, we employ the diabatization method already alluded to in Sec.~\ref{adiabatic}, where we construct the diabatic state $|\Psi_i^{(d)}(t)\rangle$ from the adiabatic basis $\left\{ |\Psi_n(t)\rangle\right\}$ using the criterion of maximal overlap with the field-free state $| \Psi_i^{(0)}\rangle$, i.e., 
\begin{align}
 |\Psi_i^{(d)}(t)\rangle &=|\Psi_n(t)\rangle, {\ \rm where\ }\hfill\\
 |\langle \Psi_n(t)| \Psi_i^{(0)}\rangle| &>| \langle \Psi_m(t)| \Psi_i^{(0)}\rangle|,\ \forall m \neq n.\nonumber
\end{align}
This can be done as long as there is one distinct adiabatic state with a prominent character of the corresponding field-free state, so that the (orthogonal) complement of adiabatic states which are mixed in is small and can be ignored. The procedure works in principle also for excited states. However, for excited states the condition of a small admixture breaks down already at low field strengths, such that this construction method works best for the field-free ground state. 
The overlap of the corresponding diabatic state $|\Psi_0^{(d)}\rangle$ with the field-free ground state $|\Psi_0^{(0)}\rangle$ is always larger than $90\%$ for field strengths considered here (see Fig.~\ref{fig3}c). Figures~\ref{fig3}(a) and \ref{fig3}(b) show the real part of the energy and the tunneling rate of $|\Psi_0^{(d)}\rangle$ as a function of the electric field. 
The shift of the real part of the energy is well approximated by a quadratic behavior; for low field strengths below $0.1$~a.u. the prefactor is in accordance with the literature value of the polarizability of the helium ground state \cite{kon, chen}. 
As expected, the tunneling rate increases considerably for sufficiently high field strengths. For field strengths larger than $0.07$~a.u. the ionization rate is well captured by the analytic expression derived in the tunneling limit of the strong-field approximation \cite{iva}.

Studying the adiabatic eigenstates and the avoided crossings reveals the suitability of the diabatic state constructed as shown above for the description of strong-field ionization. The advantage of the diabatic basis is that the system follows one single diabatic state, which gives a clear and intuitive picture for the explanation of the physics in the tunneling regime.


\subsection{Ionization dynamics}
So far, the analysis was performed for the spectrum of adiabatic eigenstates, i.e., for static electric fields. Now we introduce dynamics by 
considering a Gaussian pulse of the form
\begin{equation}
 F(t) = f(t)\,\cos(\omega t) = F_0\,e^{-t^2/2\tau ^2}\,\cos(\omega t),
\end{equation}
where $F_0$ is the peak strength of the electric field, $\tau$ is connected to the full width of the pulse at half maximum 
by $\tau^2=\rm{FWHM}^2/(8\ln 2)$, and $\omega$ is the field frequency. 

We want to calculate the ionization probability out of the diabatic state $|\Psi_0^{(d)}\rangle$ when applying this pulse. Let us assume that we have found a diabatic basis in which this particular diabatic state can be described by a coefficient $\alpha_0^{(d)}$. Then the exact wavefunction reads $\Psi(t)=\sum_{i}\alpha_i^{(d)}(t)\Psi_i^{(d)}(t)$. In analogy to the case of the adiabatic representation, equations of motion can be obtained for the coefficients in the diabatic basis where now coupling elements between the diabatic states imply nondiabatic transitions [cf. Eq.~\eqref{eom}]. If, in a ``diabatic approximation'', the nondiabatic transitions are neglected we obtain the following equation of motion for the coefficients:
\begin{equation}
 i\dot{\alpha}_i^{(d)}(t)= \left[E_i^{(d)}-i\frac{\Gamma_i^{(d)}}{2}\right]\alpha_i^{(d)}(t),\label{diabeq}
\end{equation}
where $\Gamma_i^{(d)}$ is the ionization rate of the diabatic state $i$. From the ionization rate of our distinguished diabatic state its population evolution $P_0^{(d)}(t)=|\alpha_0^{(d)}(t)|^2$ during the pulse can be inferred. To this end, the equation of motion for the probability of remaining in this particular diabatic state is calculated (we omit indices for the sake of readability):
\begin{equation}
 \frac{dP}{dt}=\frac{d}{dt}|\alpha(t)|^2=\alpha^*(t)\dot{\alpha}(t)+\dot{\alpha}^*(t)\alpha(t).
\end{equation}
Inserting Eq.~\eqref{diabeq} in this equation the following rate equation for the population is obtained (cf. Ref.~\onlinecite{lou}):
\begin{equation}
 \dot{P}(t) = -\Gamma[F(t)] \  P(t),
\end{equation}
which can be analytically solved by separation of variables:
\begin{equation}
 P(t) = \exp\left\{-\int_{-\infty}^{t} dt'\ \Gamma[F(t')]\right\}, \label{rate}
\end{equation}
with the initial condition $P(t\!=\!- \infty)\! = 1$. Note that the rate depends on the external field. Inserting the tunneling rate of the diabatic  state in Eq.~(\ref{rate}) we calculate the diabatic ionization dynamics. Thereby we observe how much is ionized out of $|\Psi_0^{(d)}\rangle$. Deviations from Eq.~(\ref{rate}) in the population dynamics can be attributed to nondiabatic behavior, i.e., transitions to other diabatic states.

\begin{figure}[htbp]
\centering
\includegraphics[width=\linewidth]{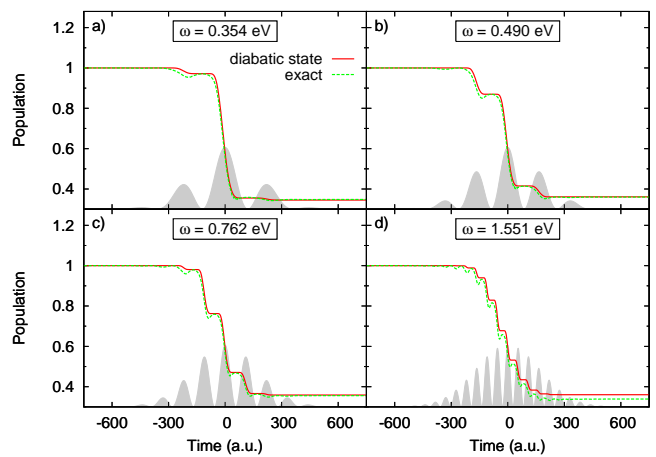}
\caption{Comparison of the ground-state populations calculated via numerical solution of the Schr\"odinger equation and via the rate equation \eqref{rate} for the distinguished diabatic state for four different photon energies. The pulse intensities are highlighted in the background: the pulse amplitude is $F_0=0.25$~a.u., and the pulse duration is $400$~a.u. ($\approx 10$~fs).}
\label{fig4}
\end{figure}

\begin{figure}
\centering
\includegraphics[width=\linewidth]{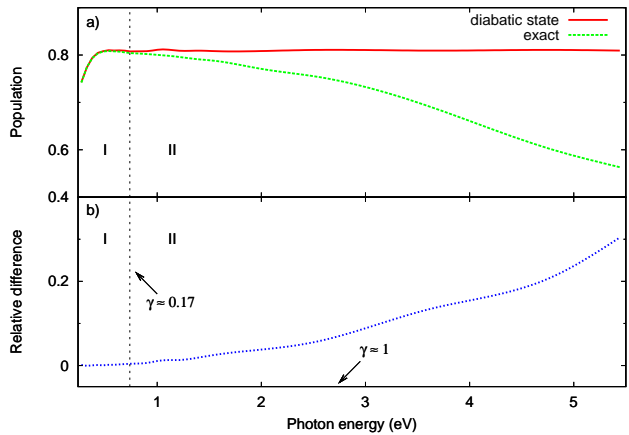}
\caption{(a) Ground-state population after the end of the pulse calculated via numerical solution of the Schr\"odinger equation 
and from the single diabatic ground state as a function of the photon energy, and (b) 
relative difference between the two results, corresponding to the degree of nondiabaticity of the ionization. 
The peak field strength is $F_0=0.2$~a.u., and the pulse duration is $400$~a.u.
The corresponding Keldysh parameter $\gamma$ is shown for different regions.}
\label{fig5}
\end{figure}

The results for four selected photon energies are shown in Fig.~\ref{fig4} 
for an electric field amplitude of $F_0=0.25$~a.u. The pulse duration is kept constant so that we can study the ionization regime from few- to multi-cycle pulses. The exact result refers to the numerical solution of the Schr\"odinger equation [see Eq.~\eqref{tdse}], where all dynamics are included, 
while the calculation of the diabatic curve via Eq.~(\ref{rate}) involves only the diabatic state $|\Psi_0^{(d)}\rangle$. 
The gray-shaded areas in the background indicate the pulse intensity.
In the frequency range shown, the evolution of the ground state population is well described by considering 
only the single diabatic state. For $\omega=0.3-0.8~{\rm eV}$ [see Fig.~\ref{fig4}a)--c)] the difference between the numerically exact 
and the diabatic calculation is insignificant, while for $\omega=1.5$~eV [see Fig.~\ref{fig4}d)] the discrepancy between the two methods becomes more noticeable.
This is exactly the difference which gives us a measure of nondiabaticity.
To clarify this further, a comparison between the two methods is shown in Fig.~\ref{fig5} 
for a peak field strength of $0.2$ a.u. by depicting the populations [Fig.~\ref{fig5}a)] and the relative difference [Fig.~\ref{fig5}b)] between them after the end of the pulse.
One can clearly see that for sufficiently low energies the total ionization probability is reproduced exactly by considering only the 
diabatic state (region I). For higher energies around $1$~eV (region II), the difference increases significantly, indicating that nondiabatic effects start to become important. 


\subsection{Nondiabaticity and the special case of few-cycle pulses}

In order to find a common way of speaking we incorporate the Keldysh parameter in our considerations, which has been used as an adiabaticity parameter.
Following our language of the adiabatic representation, the ionization in the tunneling regime, $\gamma\ll 1$, is diabatic rather than adiabatic.  
We conclude from Fig.~\ref{fig5} that in the region where $\gamma\approx 1$ the relative difference between the results calculated from the diabatic ionization rate via Eq.~\eqref{rate} and from the solution of the Schr\"odinger equation is greater than $10\%$. This is a clear sign of {\em nondiabatic} behavior. Already for $\gamma \approx 0.17$ the diabatic ionization probability starts to differ slightly from the total ionization probability. For a fixed pulse duration we can also divide the frequency range according to the number of cycles in the pulse. Starting from the highest frequencies studied here we have multi-cycle pulses, until we reach few-cycle pulses at a photon energy of $\approx0.8$~eV. 

The dynamics for few-cycle pulses is commonly considered to be nonadiabatic (in our language this translates to nondiabatic) \cite{bec,zhe}. 
We find that even for few-cycle pulses the tunneling is completely diabatic. 
In the framework of ADK theory and other approaches \cite{pon} the ionization rate $\overline{\Gamma}(t)$ 
is obtained by integrating over one period of the field \cite{bis}:
\begin{equation}
 \overline{\Gamma}(t) = \frac{1}{2\pi} \int_0^{2\pi} {d\varphi\ \Gamma[f(t) \  \cos \varphi]}, 
\label{aver}
\end{equation}
where $\Gamma[F]$ is the instantaneous ionization rate. Hence, the fact that the ADK theory of tunneling ionization and similar approaches cannot reproduce the correct (diabatic) ionization rate for few-cycle pulses is not due to coupling to higher states \cite{pot, zhe}, 
but rather because the pulse envelope changes dramatically within one cycle. In this limit the rate cannot be averaged over one period as was done 
in Eq.~(\ref{aver}), whereas for multi-cycle pulses it can be used in combination with Eq.~(\ref{rate}) yielding
\begin{equation}
 P(t) \approx \exp\left\{ -\int_{-\infty}^{t} dt'\ \overline{\Gamma}[{f}(t')]\right\}.
\end{equation}
Analyzing region I in Fig.~\ref{fig5} further, we observe that the ionization probability is not constant 
as a function of photon energy.
But the population loss in region I is well described by the ionization out of $|\Psi_0^{(d)}\rangle$. 
According to our argument above, the apparent frequency dependence is rather a dependence on the 
form of the pulse or analogously on the relation between the cycles and the pulse envelope, 
which appears in a pronounced way for few-cycle pulses. Preferably, to avoid confusion, it could be called form dependence.
As we have seen, the ionization behavior for few-cycle pulses can be well understood from the dynamics of a single diabatic state. 


\section{Conclusion}
We have studied the dynamics of tunneling ionization in atoms and have found that, within the framework of the adiabatic representation, it is diabatic rather than adiabatic.
We have identified two distinct ionization regimes depending on their diabatic behavior. 
In particular we have characterized the transition from the diabatic to the nondiabatic regime.

In the low-frequency limit the total ionization probability 
is reproduced by the contribution of the tunneling probability of one single diabatic state. 
This means that in this regime there are no significant transitions to other diabatic states. 
For few-cycle pulses, the ionization probability depends on the frequency for a fixed pulse duration. 
However, this is not a nondiabatic effect, but the effect stems from the form dependence of the pulse, 
and the consequent fact that the rate cannot be averaged any longer over one period. 

When nondiabatic transitions 
start to happen, the difference between the diabatic state ionization probability and the total probability increases dramatically. 
For frequencies in the range of the binding energy of the atom one-photon absorption can occur which is a completely nonadiabatic and even nondiabatic process. 
Already for parameters $\gamma\approx 0.17$ the diabatic ionization probability starts 
to differ noticeably from the total ionization probability, even though the perturbative multiphoton regime is not yet entered.
From the perspective of the adiabatic representation, the Keldysh parameter is found to be an approximate measure of diabaticity. 

\section{Acknowledgments}
AK is grateful to Oriol Vendrell for fruitful discussions. This work has been supported by the Deutsche Forschungsgemeinschaft 
under Grant No. SFB 925/A5.

\appendix\section{Propagation methods}\label{app}
This short section provides supplementary details on the technical features involved in our calculations. 
We need to find the exact solution of the Schr\"odinger equation [see Eq.~(\ref{tdse})] for the time-dependent Hamiltonian [see Eq.~(\ref{field})]. 
Via the propagation of the wavefunction in time, where the full dynamics are considered (adiabatic and nonadiabatic), we obtain the numerically exact solution.
In the adiabatic approach the instantaneous Hamiltonian for different field strengths (corresponding to different time steps) is diagonalized. 
In the latter case the adiabatic eigenstates are obtained, so that the adiabatic dynamics are studied. 

For both methods we employ the time-dependent configuration interaction singles (TDCIS) scheme \cite{roh, pab}. 
Starting with the Hartree-Fock ground state \cite{bet} as the field-free ground state, $|\Phi_0\rangle$, we include one-particle--one-hole configurations \cite{sza}, describing the excitations of the system. 
The TDCIS wavefunction reads:
\begin{equation}
 |\Psi(t)\rangle=\alpha_0(t)|\Phi_0(t)\rangle+\sum_{i,a}\alpha_i^a(t)|\Phi_i^a(t)\rangle,\label{tdcis}
\end{equation}
where the index $i$ symbolizes an initially occupied orbital, whereas $a$ denotes a virtual orbital to which the particle can be excited. 
This means that we consider only configurations where one particle is singly excited, thereby creating one hole. 
We use the software packages ARPACK \cite{arp} and LAPACK \cite{lap} to calculate the adiabatic eigenstates.
The two methods used permit us to compare the explicitly adiabatic dynamics with the exact calculation, where all dynamics are included.

 \bibliographystyle{apsrev}
 \bibliography{sflit.bib}

\end{document}